\documentclass[epj]{svjour}
\usepackage{graphicx}
\usepackage{bm,bbm,amsmath,amssymb}

\newcommand{\beq}{\begin{equation}}
\newcommand{\eeq}{\end{equation}}

\begin{document}
\title{Thermal diffusion and bending kinetics in nematic elastomer
cantilever}

\author{K.~K. Hon, D. Corbett and E.~M. Terentjev
}
\institute{Cavendish Laboratory, University of Cambridge, J J
Thomson Avenue, Cambridge CB3 0HE, U.K. }

\date{\today}

\abstract{  Vertically aligned mono-domain nematic liquid crystal
elastomers contract when heated. If a temperature gradient is
applied across the width of such a cantilever, inhomogeneous
strain distribution leads to bending motion. We modelled the
kinetics of thermally-induced bending in the limit of a long thin
strip and the predicted time-variation of curvature agreed
quantitatively with experimental data from samples with a range of
critical indices and nematic-isotropic transition temperatures. We
also deduced a value for the thermal diffusion coefficient of the
elastomer.
} 

\maketitle

\section{Introduction}

Liquid crystal elastomers (LCE) combine the long range
orientational correlation of liquid crystals and
entropically-driven polymer elasticity to give a range of exotic
properties such as spontaneous reversible shape changes and ``soft
elasticity" -- deformation with no or little energy cost, leading
to a variety of director instabilities under transverse extensions
\cite{MWbook}.

Nematic LCE possess the simplest uniaxial orientational order and
can be synthesized by incorporating rod-like anisotropic mesogenic
groups into the strands of weakly cross-linked polymer networks.
The order is characterized by its principal axis, the nematic
director $\bm{n}$, and the scalar order parameter $Q = \langle
P_2(\cos \theta ) \rangle$, which measures the mean orientation of
mesogenic groups with respect to the director. Such an internal
degree of freedom coupled to the elastic body constitutes what is
known as the Cosserat medium: the relative movement of
crosslinking points provides elastic strains and forces, while the
director rotation causes local torques and couple-stresses -- both
intricately connected in the overall macroscopic response of the
body. In fact, the physics of LCE is much richer than of notional
Cosserat solids because (again due to the entropic nature of long
polymer chains connecting the crosslinking points) rubbers are
capable of very large shear deformations (being at the same time
essentially incompressible). Hence, one expects a variety of
unique physical properties, especially in the region of large
deformations. However, in this work we shall explore only small
local strains.

Due to the coupling to the elastic body, the change in the degree
of alignment of mesogenic rods leads to spontaneous elongation or
contraction of the whole network along $\bm{n}$ as constituent
polymer chains become on average more or less anisotropic (prolate
or oblate depending on the system). This direct coupling between
physical conformation and order parameter has been theoretically
predicted a long time ago \cite{Gelling}, and then comprehensively
demonstrated by simultaneous measurements of length and order
using diffraction techniques \cite{Kaufhold,Kupfer2,Tajbakhsh}.

Landau theory predicts a 1st order transition into the isotropic
phase as the nematic LCE is heated above its nematic-isotropic
transition temperature $T_{NI}$. This is based on the quadrupolar
symmetry of the second-rank tensor order parameter of the nematic,
which does not distinguish between rods pointing ``upwards" and
``downwards". However, in the elastomer network that was
crosslinked in the aligned director state in order to obtain a
permanent monodomain nematic texture \cite{Kupfer1}, one does not
find a discontinuous jump in the order parameter. Instead,
frozen-in uniaxial stress leads to the supercritical continuous
change of $Q(T)$ across the transition, and with it -- the
continuous equilibrium uniaxial deformation of the monodomain
nematic LCE. Depending on the degree of induced anisotropy of
polymer chains forming the network, the magnitude of this
deformation can be as high as 500\% \cite{Ahir1}.

Spontaneous shape changes can also be induced optically.
Photoelastomers doped with rod-like groups, such as azobenzene
derivatives, which undergo {\em trans-cis} isomerization on
absorption of UV photons \cite{Nishi,Hogan}, or carbon nanotubes,
which respond to IR light \cite{Ahir2}, are found to contract when
irradiated at suitable wavelengths since local order is disrupted
by the kinked dopant groups. Due to the high stroke and the
equilibrium (reversible) nature of induced deformations, this now
becomes an active area of engineering micro-optical mechanical
systems (MOMS).

Inhomogeneous deformations are of special interest since they see
potential applications in photo- and thermal actuators, detectors
and sensors, microrheological valves and pumps, as well as
structures which can respond to their neighboring environment.
Non-uniform deformations occur when a spatial stress distribution
is induced inside an elastomer. This could be achieved by
irradiation, or by application of temperature gradient across the
sample. Mathematical models \cite{Maha} have been proposed to
predict equilibrium curvatures in unilaterally illuminated
photoelastomer cantilevers with exponential attenuation. However,
in contrast to uniaxial contractions along the nematic director
that have been well documented, no quantitative measurements have
been made so far on bending curvatures. At the same time, while
qualitative experiments on optically-induced deformations have
reported time-scales varying from $<100$ms \cite{Palffy} to  $\sim
1-10$s \cite{Ikeda,nelson} depending on incident intensity of
light sources, the kinetic aspects of the bending motion have not
been addressed theoretically.

This paper presents the first quantitative experimental study of
the dynamics and  kinetics of thermally induced bending in a
nematic elastomer cantilever. We apply radiative heating to one
side of cantilevers made from well-aligned monodomain polysiloxane
side-chain elastomers, and measure the amplitude and time
evolution of the induced curvature. We also develop a theoretical
model, which predicts the reduced curvature of the cantilever as a
function of time for cantilevers with with different critical
exponents, transition temperature and maximum strain. A value for
the thermal diffusion coefficient of the elastomer is estimated
from matching the model predictions to the experiment.

\section{Experimental section}

\begin{figure} 
\centering
\resizebox{0.47\textwidth}{!}{\includegraphics{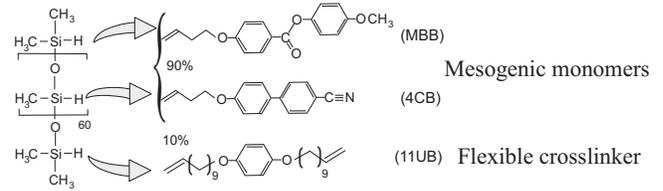}}
\caption{Schematic illustration of the materials used in this
work. Siloxane backbone chain with Si-H groups was reacting with
90 mol\% mesogenic side groups and 10 mol\% of flexible
crosslinking groups (11UB). Two materials differed in the
composition of mesogenic groups: NE-A had 90mol\% of MBB, while
NE-B had 70mol\% of MBB and 20mol\% of ACB. }
 \label{chemistry}
\end{figure}

\emph{Materials.} \ All side-chain siloxane liquid crystalline
elastomers, as well as their starting materials, were prepared in
the Cavendish Laboratory following the procedures of Finkelmann et
al. \cite{Kupfer1,Greve}. The polymer backbone was a
poly-dimethylhydrosiloxane with approximately 60 Si-H units per
chain, obtained from ACROS Chemicals. The pendant mesogenic group
in sample A (NE-A) was purely 4-methoxyphenyl-4-(1-buteneoxy)
benzoate (MBB), while sample B (NE-B) contained of 70mol\% of MBB
and 20mol\% of 4-alkeneoxy-4'-cyanobiphenyl (ACB), as illustrated
in Fig.~\ref{chemistry}. All networks were chemically crosslinked
via the same hydrosilation reaction in the presence of commercial
platinum catalyst COD, obtained from Wacker Chemie, with
di-functional crosslinking group 1,4 di(11-undecene) benzene
(11UB) also synthesized in-house. In all cases the crosslinking
density was 10 mol\% of the reacting bonds in the siloxane
backbone, so that on average each chain has 9 mesogenic groups
between crosslinking sites. These materials are very well studied
over the years; both have a glass transition around 0C and
nematic-isotropic transitions: $T_{NI} \approx 87$C for NE-A, and
$T_{NI} \approx 101$C for NE-B.

\emph{Monodomain alignment.} \ Mono-domain, aligned samples of
nematic elastomers were made by following the classical two-step
crosslinking approach of Finkelmann et al. \cite{Kupfer1}. First
we prepare partially crosslinked films in a centrifuge, highly
swollen in toluene (2-3ml per 1g of material), reacting for 25-35
minutes at $\sim$75C before evaporating the solvent and suspending
the samples under load in an oven for more than 5 hours at 120ºC
to complete the second-stage crosslinking reaction. A careful
study of reaction kinetics ensured that approximately 50\% of
crosslinks were established in the first stage of this
preparation. When a uniaxial stress is applied to such a partially
crosslinked network, the uniaxially aligned state in the resulting
nematic elastomer is established with the director along the
stress axis. This orientation is then fixed by the subsequent
second-stage reaction, when the remaining crosslinks are
established.

Following the original ideas of \cite{Kupfer1} and the present
understanding of the nature of polydomain nematic LCE
\cite{Fridrikh}, in all cases we performed the second stage
crosslinking in the high-temperature isotropic phase: only in this
way a good alignment and mechanical softness are achieved (in
contrast to crosslinking in a stretched polydomain nematic phase,
which results in topological defects and localized domain walls
frozen in the material).

The mechanical history of the samples was eliminated by annealing
in the isotropic phase for $>2$ hours ($\sim 130^{\rm o}$C)
followed by slow cooling. Precise measurement of variation in
natural length $L(T)$ with temperature was then made with a
travelling microscope, which followed the end points of a sample
that was suspended without load and heated at a slow rate of
0.33°C/min in an insulated glass-front oven.

\begin{figure} 
\centering
\resizebox{0.32\textwidth}{!}{\includegraphics{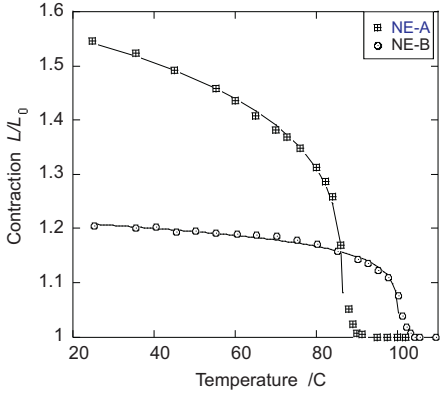}}
\caption{Curves of equilibrium uniaxial contraction of monodomain
nematic LCE. The two samples (labelled on the plot) have different
transition temperatures $T_{NI}$ and different chain anisotropy,
leading to the 20\% and 55\% contraction respectively. Solid lines
drawn through the data below $T_{NI}$ represent the analytical fit
functions described in the text.} \label{contraction}
\end{figure}

Measurements of natural length $L(T)$ variation with temperature,
Fig.~\ref{contraction}, were fitted to a model function $L/L_0 = 1
+ \beta (1- T/T_{NI})^a$, where $L_0$ is the constant length of
samples in the isotropic phase. Obviously, such a superficially
critical behavior cannot be matched to experiment at the
transition point itself (where supercritical effects take over),
but it provides a very good continuous interpolation of the data
in the nematic phase. Fitting to the data gives $a=0.25, \, \beta
=0.843$ and $T_{NI}=359.6$K for NE-A,  and  $a=0.21, \, \beta =
0.3$ and $T_{NI}=373.8$K for NE-B.

The thickness $w$ of the two samples (which we shall require in
the cantilever analysis) was 0.351mm (NE-A) and 0.368mm (NE-B) at
room temperature. Elastomer samples were cut into thin strips of
approximate dimensions $L \times W \times w =$ 5mm $\times$ 1mm
$\times w$ (with differing thickness) and had one end vertically
attached to a stand on an adjustable platform. Sideways images of
the strips (cantilevers) were taken by a Sanyo VCB-3512T
monochrome CCD camera (f = 9mm) with direct back lighting and
digitally captured using software FTA32 by First Ten Angstroms Inc
at a frame rate of 15fps. An Antex CS 16W soldering iron provided
heating. This soldering iron had a flat tip (cylinder of
4.5mm-diameter) which was providing uniform radiative heating over
the whole cantilever. Imaging of this tip also acted as a scale
for confirming the thickness $w$ of the samples by comparing
dimensions on-screen. The soldering iron, which was allowed to
equilibrate for 15 minutes before each experiment, was mounted
horizontally on a movable stand which can be slid to the desired
position (~2mm) in front of the mounted sample in under 0.2s,
which marked the start of each kinetic measurement. Temperatures
at the front and the back of the samples were measured with a
thermocouple, however, not during the cantilever-bending
experiment (but in a separate event of heating in exactly the same
conditions).

\begin{figure} 
\centering
\resizebox{0.4\textwidth}{!}{\includegraphics{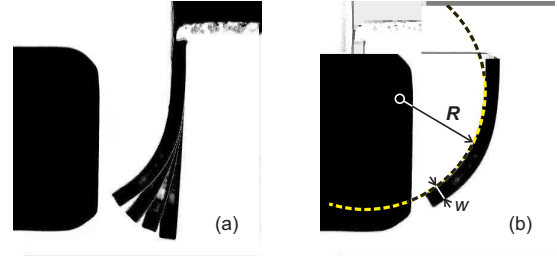}}
\caption{(a) A composite image showing the extent of cantilever
bending. On the left one can see the tip of soldering iron. \ (b)
The scheme of manual analysis of cantilever curvature. }
\label{photo}
\end{figure}

Movies of the bending motion were taken and each frame was
analyzed both manually and using a MATLAB image-processing
algorithm. Manually, the radius of curvature was obtained by
superposing circles of various sizes on the outline of the curved
sample using graphics software CorelDraw and adjusting until the
circle of best fit was found, see Fig.~\ref{photo}. The automated
MATLAB algorithm extracted the position of points along the curved
edge of the sample in the image and fitted the set of points to
the equation of a circle with variable radius $R$ by a
least-squares method; the optimized value of $R$ was then output
as the radius of curvature. In the end, our procedure was to
analyze all images in an automated way, but then re-examine every
anomalous point manually (because we found that our algorithm was
not coping well with the cases of non-uniform curvature along the
cantilever). The outputs of this analysis were the values of
normalized curvature $w/R$ against time for each bending
experiment.

\section{Theoretical model}

\noindent {\bf Heat diffusion across a flat strip}

The problem of thermal diffusion in a flat sample exposed to a
constant heat flux from one side is certainly a classical one. We
give its brief account here in order to expose the key parameters
of the problem, required for the subsequent description of
cantilever bending. Consider a 1D diffusion of the scalar
temperature field $T(x,t)$ across the thickness of the cantilever,
$ \partial _t T = D \partial^2_x T$, where the diffusion
coefficient $D=\kappa /C$ is the ratio of thermal conductivity
$\kappa$ (given by the heat flux definition $J = -\kappa
\partial_x T$) to specific heat capacity per unit volume $C$. We
assume that the soldering iron acts as a source of constant flux
$J$ at $x=0$ while losses on the front ($x=0$) and back ($x=w$)
surfaces are taken to be proportional to the difference between
the temperature on the sides, $T_{\rm f}$ and  $T_{\rm b}$, and
the ambient surrounding, $T_0$, i.e. $- \kappa \partial_x T = J -
\gamma (T_{\rm f} - T_0)$ at $x=0$,  and $- \kappa \partial_x T =
\gamma (T_{\rm b} - T_0)$ at $x=w$, taking flux to be positive to
the right.

Since the diffusion equation consists only of derivatives of the
temperature field T and the boundary conditions are only sensitive
to temperature differences, we can homogenize the problem by
considering the function $\theta = (T-T_0)$ instead. Introducing
natural variables $\chi = x/w$ and $\tau = Dt/w^2$, the problem
can be recast as:
\begin{equation}
\partial_\tau \theta = \partial^2_\chi \theta \ , \ \ {\rm with}
\ \left\{
\begin{array}{ll}
\partial_{\chi} \theta &= \Delta (\theta - \Theta_c) \ {\rm at} \ \chi=0 \cr
\partial_{\chi} \theta &= -\Delta \, \theta  \ \ \  {\rm at} \ \chi=1 \cr
\theta &= 0 \qquad {\rm at} \ \tau=0
\end{array}  \right.
\end{equation}
where $\Delta = w \gamma/\kappa$ and $\Theta_c = J/\gamma$ are the
two essential parameters of the problem.

In the steady state $\partial_\tau \theta = 0$. It being a 1-D
problem, only a linear solution $\theta = A \chi + B$ could
satisfy $\partial^2_\chi \theta=0$. Letting the steady state front
and back temperatures be and $T_{\rm f}^*$ and $T_{\rm b}^*$
respectively (equivalently $\theta_{\rm f}^*$ and $\theta_{\rm
b}^*$), the steady-state temperature profile across the sample is
given by $\theta_{\rm s} = \theta_{\rm f}^* - (\theta_{\rm f}^* -
\theta_{\rm b}^*) \chi$. Parameters $\Delta$ and $\Theta_c$ can
then be expressed as
 \begin{equation}
\Delta = (\theta_{\rm f}^*/\theta_{\rm b}^* -1) \ \ {\rm and} \ \
\Theta_c = (\theta_{\rm f}^* + \theta_{\rm b}^*),
 \end{equation}
which consist only of explicitly measurable quantities $T_{\rm
f}$,  $T_{\rm b}$ and $T_0$. Note that for a thin enough sample
one expects $T_{\rm f}^* \approx T_{\rm b}^*$ and so $\Delta \ll
1$. The full time-dependent solution can be obtained by
superposing a series of different time-decay modes $\theta = A_k
\sin(k \chi + \phi) \exp (-k^2 \tau)$ on the steady state
solution. Quantization conditions for $k$ and $\phi$ are
determined by the initial/boundary conditions, giving the
transcendental equations
 \begin{equation}
 k_n =  \Delta \tan ( n \pi/2 -k_n/2) \ \ {\rm and}
 \ \ \phi_n = ( n \pi/2 -k_n/2)  \label{kn}
 \end{equation}
for integer $n$. Fourier analysis of orthogonal modes in $\theta
(\chi,\tau)$ gives the expression for coefficients
\begin{eqnarray}
A_n &=&    (-1)^{(n+1)/2} \frac{2 \Theta_c \sin(k_n/2)} {k_n +
\sin(k_n)} \ \ \ {\rm  for \ odd} \ n \ ; \\
  &=& (-1)^{n/2 +1} \frac{2 \Theta_c}{k_n} \frac{\Delta}{2+\Delta}
  \frac{k_n \cos(k_n/2) - 2 \sin(k_n/2)}{k_n - \sin(k_n)}  \nonumber
\end{eqnarray}
for  even $n$. The full solution for the temperature across the
sample of thickness $w$ is therefore given by
 \begin{eqnarray}
T(x,t) &=& T_{\rm f}^* - (T_{\rm f}^*-T_{\rm b}^*) \frac{x}{w} \label{finalT}
\\
&& + \sum_{n=1}^\infty A_n \sin \left(k_n \left[ \frac{x}{w} -
\frac{1}{2} \right] + \frac{n \pi}{2} \right) e^{-k_n^2 D t/w^2}
\nonumber
 \end{eqnarray}
There are two features of this solution that we need for our main
problem. First of all, there is a characteristic time scale in the
problem, given by the ratio $w^2/D$ which will allow us estimate
the thermal diffusion constant in nematic LCE. Note that the
$(n=1)$ mode in Eq.(\ref{finalT}) has $k_1 \approx \pi \Delta$ at
$\Delta \ll 1$ and, therefore, this is the slow-decaying mode.
Other modes have $k_n$ of order $(n-1)\pi$ and decay fast, in
practice, within a few seconds in our experiments.

\begin{figure} 
\centering
\resizebox{0.4\textwidth}{!}{\includegraphics{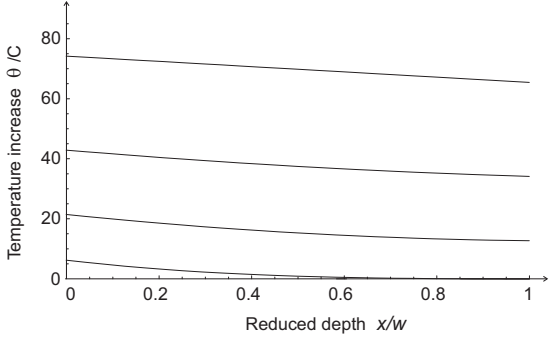}}
\caption{Plots of $\theta = T-T_0$ across the cantilever thickness
$x/w$. Increasing curves are for the times $D t/w^2 = 0.1, \, 1,
\, 3$ and $100$.  } \label{Tx}
\end{figure}

The second aspect of the solution $T(x,t)$ is the rather smooth
variation across the cantilever thickness. Figure~\ref{Tx}
demonstrates the $x$-dependence at different times, which
justifies an essential simplifying approximation made in the next
section, taking $T(x)$ to be a simple linear function connecting
the two
values $T_{\rm f}$ and $T_{\rm b}$. \\

\noindent {\bf Kinetics of cantilever bending}

Consider now a long thin strip of elastomer, preferentially
contracted at the front and bent accordingly, due to unilateral
heating from the side $x=0$ starting at $t=0$. A temperature
distribution $T(x,t)$ is set up across the cantilever thickness.
Since the length of a mono-domain nematic elastomer below $T_{NI}$
is locally given by $L/L_0 = 1 + \beta (1- T/T_{NI})^a$, the local
strain distribution due to contraction along the $z$-direction
(vertical in Fig.~\ref{photo}) can be calculated. Taking the
zero-strain state at ambient temperature $T_0$, we obtain:
 \begin{eqnarray}
 \varepsilon (x) &=& \frac{L(x)}{L(T_0)} -1  \\
 &=&  \frac{\beta}{ [ 1 + \beta (1 - T_0/T_{NI})^a ] } \left[ \left( 1-
\frac{T}{T_{NI}} \right)^a - 1\right] \, ,  \nonumber
 \end{eqnarray}
 where $L(T_0)$ is the sample length at ambient temperature.
Assuming Young's modulus to be constant over the temperature and
strain range of the experiment, this strain can be directly
converted to stress $\sigma = E \, \varepsilon (x)$, at each depth
$x$ into the sample. As the elastomer deforms incompressibly, a
contraction in the $z$-direction would lead to transverse
expansions along $x$ and $y$. However, for the case of a long thin
strip in which $L \gg W
> w$, $x$- and $y$- curvatures can be safely ignored.

Let $x_n$ be the position of a neutral plane \cite{Maha}.
Mechanical equilibrium requires, in the absence of external forces
and torques, that force and moment vanish across every
cross-section of the cantilever. This means that all bending
stress, $E(x-x_n)/R$ where $R$ is the radius of curvature of the
beam, is provided by the excess stress $E[ \varepsilon (x)-
\varepsilon (x_n)]$ with respect to this neutral plane. Two
conditions representing the balance of forces and torques are, as
in \cite{Maha},
\begin{eqnarray}
\int_0^w W\, dx \, E \frac{x-x_n}{R} &=& \int_0^w W\, dx \, E[
\varepsilon (x)- \varepsilon (x_n)] \\
\int_0^w W\, dx \, x \, E \frac{x-x_n}{R} &=& \int_0^w W\, dx \,
x\, E[ \varepsilon (x)- \varepsilon (x_n)] \nonumber
\end{eqnarray}
where $W$ is the width of the cantilever (the $y$-dimension), $Ww$
being the cross-section area of the beam.

Assuming a linear temperature drop across the thickness,
$T(x)=T_{\rm f} - (T_{\rm f} - T_{\rm b}) x/w$, the spatial
integrals can then be conveniently converted to over temperature
via $dT = - (T_{\rm f} - T_{\rm b}) dx/w$ to obtain
\begin{eqnarray}
\frac{1}{R} \left( \frac{w}{2} - x_n \right) &=& \int_{T_{\rm
b}}^{T_{\rm f}} \frac{ \varepsilon (T) \, dT}{T_{\rm f}-T_{\rm b}}
- \varepsilon (x_n)
 \\
\frac{1}{R} \left( \frac{w}{3} - \frac{x_n}{2} \right) &=&
\int_{T_{\rm b}}^{T_{\rm f}} \frac{(T_{\rm f}-T) \varepsilon (T)
\, dT}{(T_{\rm f}-T_{\rm b})^2} - \frac{\varepsilon (x_n)}{2} .
\nonumber
\end{eqnarray}
Note that both the lateral width $W$ and the Young modulus $E$
scale out of these mechanical balance equations. On elimination of
$x_n$ and $\varepsilon (x_n)$ from these two equations, one
obtains
\begin{equation}
\frac{w}{R} = 12\int_{T_{\rm b}}^{T_{\rm f}} \frac{(T_{\rm f}-T)
\varepsilon (T) \, dT}{(T_{\rm f}-T_{\rm b})^2} - 6 \int_{T_{\rm
b}}^{T_{\rm f}} \frac{\varepsilon (T) \, dT}{T_{\rm f}-T_{\rm b}}
\end{equation}
The problem could be solved completely if we had an analytical
expression for $\varepsilon (T)$. Unfortunately, our interpolation
formula is only applicable below $T_{NI}$, while above the
transition $\varepsilon$ is constant. The break at $T_{NI}$ calls
for a different mathematical treatment for the following two
regimes:

(i) When the temperature is below $T_{NI}$ everywhere in the
sample, $T_{NI} > T_{\rm f} > T_{\rm b}$, evaluating the integrals
gives the dimensionless curvature:
 \begin{eqnarray}
 \frac{w}{R} &=& \frac{6\beta}{ [ 1 + \beta (1 - T_0/T_{NI})^a ] }
  \frac{1}{2+3a+a^2}
  \cdot \label{w1} \\
 && \ \   \left[
\frac{T_{NI}-T_{\rm f}}{T_{\rm f}-T_{\rm b}}\left(1-\frac{T_{\rm
f}}{T_{NI}} \right)^a    \left(2+a+2\frac{T_{NI}-T_{\rm f}}{T_{\rm
f}-T_{\rm b}} \right)  \right.  \nonumber  \\
 && \qquad + \left. \frac{T_{NI}-T_{\rm
b}}{T_{\rm f}-T_{\rm b}}\left(1-\frac{T_{\rm b}}{T_{NI}} \right)^a
\left(a-2\frac{T_{NI}-T_{\rm f}}{T_{\rm f}-T_{\rm b}} \right)
\right] \nonumber
\end{eqnarray}
Here $T_{\rm f}$ and $T_{\rm b}$ are independently measurable
time-dependent functions. At all temperatures within this regime
the curvature $w/R$ is a monotonically increasing function of time
with the positive second derivative (convex function).

(ii) When the $T_{\rm f}$ exceeds $T_{NI}$, so that the phase
transition front is inside the sample, $T_{\rm f} > T_{NI} >
T_{\rm b}$, we obtain:
 \begin{eqnarray}
 \frac{w}{R} &=& \frac{6\beta}{ [ 1 + \beta (1 - T_0/T_{NI})^a ] }
\frac{T_{NI}-T_{\rm b}}{T_{\rm f}-T_{\rm b}} \cdot \label{w2}
\\ && \ \ \left(1-\frac{T_{\rm b}}{T_{NI}} \right)^a
\frac{a+2(T_{\rm f}-T_{NI})/(T_{\rm f}-T_{\rm b})}{2+3a+a^2}
\nonumber
\end{eqnarray}
This expression represents a non-monotonic function of time, with
the negative second derivative (concave function) and the maximum
curvature $w/R$ followed by a decrease when the majority of the
sample becomes isotropic.

The regime (ii) comes to an end when the temperature at the back
of the cantilever, $T_{\rm b}$ reaches the transition point
$T_{NI}$, i.e. all of the sample becomes homogeneously isotropic.
At this point, evidently, $w/R =0$. Calculating the integrals in
the regime when all of the sample is isotropic, $T_{\rm f} >
T_{\rm b} > T_{NI} $, confirms that $w/R=0$ at all times.

In our experiments the powerful heating flux has ensured that the
temperature rise was high  so the interesting regimes (i) and (ii)
occurred at relatively short times, when both $T_{\rm f}$ and
$T_{\rm b}$ were well-approximated by a single exponential:
\begin{eqnarray}
T_{\rm f} &=& T_{\rm f}^* - (T_{\rm f}^*- T_0) e^{-t/\tau_{\rm f}}
\label{tauFB}
\\ T_{\rm b} &=& T_{\rm b}^* - (T_{\rm b}^*- T_0) e^{-t/\tau_{\rm
b}}\nonumber
\end{eqnarray}
where $\tau_{\rm f}$ and $\tau_{\rm b}$ are the effective thermal
diffusion times at the front and back, respectively. It is
expected that $\tau_{\rm f}$ and $\tau_{\rm b}$ would take similar
but not identical values, since inspection of the full series for
$T(x,t)$ shows that the spatial coefficient of $\sin[k_n(x/w -
1/2) + n\pi/2]$ is identical at $x=0$ and $x=w$ for odd modes, but
swaps sign for even $n$. In this single-exponential approximation,
regime changes occur at $t_1 =  \tau_{\rm f} \ln (T_{\rm f}^*-
T_0)/(T_{\rm f}^*- T_{NI})$, when the front of the elastomer
enters the isotropic phase and the curve $w/R(t)$ has an
inflection point; and at $t_2 = \tau_{\rm b} \ln (T_{\rm b}^*-
T_0)/(T_{\rm b}^*- T_{NI})$, when the elastomer becomes uniformly
isotropic and hence returns to a state of zero curvature, with
also a zero tangent. The values and gradients of reduced curvature
match on both sides of $t_1$ and $t_2$, as is physically required.

\section{Analysis of bending kinetics}

In both samples, the reduced curvature $w/R$ variation is
characterized by three distinct regimes. Immediately after the
start of the experiment (heat flux on) there is a slow initial
increase of curvature with time; this is followed by a sharp peak,
after which the curvature rapidly dropped to zero. This trend
corresponds to the regimes (i) and (ii) described in the
theoretical model, but also could be qualitatively understood by
considering the shape of the contraction curve of the nematic
elastomer, $L/L_0$ against $T$ (see Fig.~\ref{contraction}).
Initially, when the sample is relatively far away from $T_{NI}$,
the mechanical response to temperature change is flat. The local
strains induced at the front and back surfaces are small and
similar in magnitude. However, as the temperature approaches
$T_{NI}$, which happens first on the front of the cantilever, the
internal stress gradient is amplified by the increasing steepness
of $L(T)$ curve. This leads to cantilever curvature rising at an
increasing rate. Curvature will be decreasing again after the
front portion of the cantilever turns isotropic and stops
contracting while the on-going contraction at the back reduces the
stress gradient. Finally, the sample returns to a stationary
unbent state as it becomes uniformly isotropic.

\begin{figure} 
\centering
\resizebox{0.35\textwidth}{!}{\includegraphics{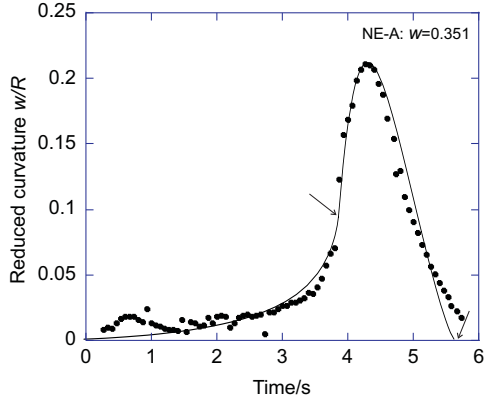}}
\caption{Experimentally measured values of reduced curvature
$w/R$, for the NE-A cantilever, against time after the start of
heating. The solid line is the fit by theoretical equations
(\ref{w1}) and (\ref{w2}), with arrows showing where the regimes
(i) and (ii) end. } \label{NEA}
\end{figure}

\begin{figure} 
\centering
\resizebox{0.35\textwidth}{!}{\includegraphics{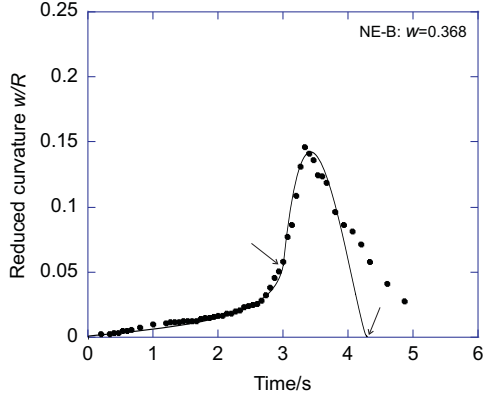}}
\caption{Experimentally measured values of reduced curvature
$w/R$, for the NE-B cantilever, against time after the start of
heating. The solid line is the fit by theoretical equations
(\ref{w1}) and (\ref{w2}), with arrows showing where the regimes
(i) and (ii) end. } \label{NEB}
\end{figure}

Figures~\ref{NEA} and \ref{NEB} show the measured values of
curvature $w/R$. The solid line in each plot is the fit by the
theoretical model, Eqs.(\ref{w1}) and (\ref{w2}). Evidently, the
agreement is very good, both qualitatively and quantitatively --
except in the final stages of sample un-bending, where the
discrepancies are significant. This, however, should be expected
because of the following two factors, one practical, the other to
do with data analysis.

In our model, we have ignored the fact that, as curvature
increases, the lower end of the cantilever lifts and becomes
closer to the heat source. Figure~\ref{photo} shows this very
clearly. This introduces a significant deviation from the
theoretical assumption in the model, that the heat flux $J$ is
constant. It is clear that, after the point of maximal bending is
reached, the real heat flux on the sample is inhomogeneous along
the length of the cantilever ($z$- local coordinate). As a result,
the far end of the cantilever (the part closest to the heat
source) will become homogeneously isotropic much earlier than the
simple 1D diffusion model would predict. In practice we see this
very clearly, as in the final stages of heating cycle the far end
of the cantilever is already straight, while the middle and near
parts still have curvature remaining. Such buckling occurred more
significantly in NE-B than in NE-A, as seen from the curvature
plots, however in all cases both manual and algorithmic fitting
was difficult and ambiguous as curvatures are no longer constant.

The other discrepancy factor is inherent in the model itself. In
order to obtain closed-form expressions for $w/R(t)$,
Eqs.(\ref{w1}) and (\ref{w2}), we had to use an interpolating
formula for the underlying thermal contraction $L(T)$. As it is
clear from Fig.~\ref{contraction}, this formula does not work in
the immediate vicinity of the transition $T_{NI}$, where all
materials show differing degree of diffuse supercritical behavior.
So it is not surprising that the model expressions deviate from
the actual data when the temperatures of the sample are around, or
slightly above, the notional $T_{NI}$. These are the regions past
the curvature peak, where the model ``assumes'' the local regions
with $T>T_{NI}$ are fully isotropic, while in practice we know the
contraction continues for 2-3 more degrees.

Among other ignored effects, which might become apparent at high
temperatures (later times of the bending cycle) there is heating
of surrounding air and lateral heat loss, which would require
ambient temperature $T_0$ to be time-dependent and a full 3-D
treatment of heat diffusion respectively. All these factors could
be eliminated or much reduced in impact. We could (and indeed have
in some experiments) mount the heat source at an angle to minimize
the effect of one cantilever end approaching it too close. We
could also write a much more elaborate interpolation formula to
account for the full continuous $L(T)$ variation, and then proceed
to calculate all integrals numerically. However, on reflection we
have decided that the benefits of such improvements would not be
worth the price of losing the simplicity of experiment and the
ease of analysis. After all, the agreement of the model with
experiment in the early regions of the bending process is
excellent, as we expect when the material is in the nematic phase.

The point of maximum curvature occurred at 4.3s for NE-A and at
3.3s for NE-B. Since we have measured the saturation temperatures
$T_{\rm f}^*$ and $T_{\rm b}^*$ independently, as well as
determined the parameters $\beta, a, T_{NI}$ of the intrinsic
thermal contraction curves for each material, the only two fitting
parameters are the front- and back-relaxation times $\tau_{\rm f}$
and $\tau_{\rm b}$, cf. Eq.(\ref{tauFB}). The best fits in
Figs.~\ref{NEA} and \ref{NEB} were achieved with $\tau_{\rm f}=$
2.42s and $\tau_{\rm b}=$2.05s for NE-A, and with $\tau_{\rm
f}=$2.13s and $\tau_{\rm b}=$2.5s for NE-B. It must be remarked
that despite the experimental and theoretical difficulties
discussed above, the two-stage increase in the cantilever
curvature shows good agreement with the experimental data. This
proves that we understand the underlying physics of
 nematic elastomer cantilevers correctly, and allows us to extract
 relevant material parameters from the fits.

Parameters $\tau_{\rm f}$ and $\tau_{\rm b}$ can be viewed as a
characteristic heating time for the sample. By comparison with the
first terms of the thermal diffusion solution, Eq.(\ref{finalT}),
they are expected to be of order $w^2/k_1^2D$. Steady state
temperatures $T_{\rm f}^*$ and $T_{\rm b}^*$ were measured to be
375K and 365K for NE-A, and 398K and 390K for NE-B, giving $\Delta
= 0.89$ and $0.133$ respectively. The value of $k_1$ can then be
obtained from solving the transcendental Eq.(\ref{kn}) numerically
(giving $k_1=0.419$ for NE-A and $0.511$ for NE-B). Rearranging,
the thermal diffusion coefficient is given by $D \sim
w^2/k_1^2\tau$. A typical value of $D$ is therefore estimated to
be $\sim 1.5 \cdot 10^{-7} \hbox{m}^2/\hbox{s}$. We are not aware
of any measurements of thermal diffusion in nematic LCE, but this
estimate compares favorably with the literature values of $D = 1.1
\cdot 10^{-7} \hbox{m}^2/\hbox{s}$ for a crosslinked silicone
elastomer \cite{rubber},  a value also consistent with
uncrosslinked silicone polymer melts \cite{silicone}.

\section{Conclusions}

In this work we experimentally studied the time-variation of
curvature of a long thin strip of aligned monodomain nematic
elastomer,  for two samples differing in their transition
(constitutive) behavior and shape dimensions.  Associated
theoretical analysis was able to quantitatively describe the data
and reflect all characteristic trends. Fitting the data allowed us
to deduce characteristic time scales and estimate the thermal
diffusion constant $D$ of the siloxane elastomer.

To the best of our knowledge, this is the first quantitative study
of thermal diffusion in nematic elastomers, as well as their
cantilever bending due to induced inhomogeneous strains arising
from unilateral radiative heating. Such controlled and
reproducible bending is an important physical effect underlying
many engineering applications. Perhaps more practically relevant
is the photo-induced cantilever bending, where the local strains
are induced due to photoisomerization reaction in azobenzene
derivatives \cite{Maha,Palffy,Ikeda,nelson,Harvey}. Our
experiments on photo-bending, analogous to the current work, will
be reported elsewhere. Nevertheless, thermal bending is a
fundamentally important effect where one tests the details of
continuum mechanics, kinetics of local and global response, and
the general understanding of nematic elastomer state.

\subsection*{Acknowledgments} We wish to thank A.R. Tajbakhsh for
preparation of samples, M. Warner for useful discussions
concerning the diffusion equation and C. Picard for assistance in
Matlab programming.


\end{document}